\documentstyle[12pt]{article}

\title{Decay on several sorts of heterogeneous centers:
Special monodisperse approximation in the situation
of strong unsymmetry. 1. General results }

\author{V.Kurasov}

\date{Victor.Kurasov@pobox.spbu.ru}

\begin{document}

\maketitle

\section{Introduction}

Metastable phase decay on the several types  of heterogeneous centers
remains a rather actual problem for theoretical investigation. For the first
time the theory for the kinetics description was
constructed
in \cite{Multidecay}. This approach  decomposes the general situation
into  characteristic situations  which  are  rather  simple.  All
limit
situations characterized by small values of characteristic parameters
can be solved by the slightly modified versions of the iteration method
initially
proposed in \cite{Kunidec}. Only the intermediate situation requires a
special method of consideration which is based on the special monodisperse
approximation.

When the total number of the heterogeneous centers is one and the same for
different types of centers then as it is shown in \cite{over} there are only
two characteristic situations: the intermediate situation and the situation
of the strong unsymmetry.

The special monodisperse  approximation is well based  and  can
be spread on a more general situation.   This generalization is important
not only to present the description in a more compact form. The special
monodisperse approximation
allows to reduce the error appeared in the limit situations. Really, in
the situation of the strong unsymmetry (this is a standard limit situation)
one has to use the monodisperse approximation. If we use the special
monodisperse
approximation instead of the already used one we shall reduce the error.

The universal character of obtained solution was shown in \cite{book1}.

The general recipe to use the special monodisperse approximation  was
suggested in \cite{book2} in the abstract manner. So, now it rather natural
to show how the special monodisperse approximation works in the limit
situations concretely.

This publication is intended to show how to use the special monodisperse
approximation in the limit situations. We shall use for example the situation
of the strong unsymmetry which appeared when the number of the first type
centers equals to the
number of the second type
centers (see \cite{over}). This situation together with the
intermediate
situation completes the consideration of the general case \cite{over}.

\section{Formulation of the problem}

Consider the system with two sorts of heterogeneous centers (they are
marked by subscripts $1$ and $2$). Suppose that
the total numbers of
heterogeneous centers $\eta_{tot1}$ and $\eta_{tot2}$ are equal
\begin{equation} \label{oo}
\eta_{tot1} = \eta_{tot2}
\end{equation}

At the initial moment of time which is denoted by the subscript $*$
there exists only heterogeneous centers. We shall call the period of
intensive formation of the droplets on the heterogeneous centers of given
type as the nucleation on the centers of given type.

We shall use in the estimates some characteristic values. Denote by $\Delta_1
t$ the duration of the nucleation on the first type centers and by $\Delta_2
t$ the duration of the nucleation on the second type centers.

Consider the situation when the  rate of nucleation on the first type
of heterogeneous centers strongly exceeds the rate of nucleation on the
second type of heterogeneous centers.
We shall choose the sorts of heterogeneous centers to have
$$
f_{*1} \gg f_{*2}
$$
where
$f_{* i} $ is the amplitude value of the stationary distribution function.

Due to (\ref{oo}) the last inequality is practically equivalent to
$$
\exp(-\Delta F_1)  \gg \exp(-\Delta F_2 )
$$
where $\Delta F$ is the height of the activation barrier.

The power of
metastability is characterized by the supersaturation
$$
\zeta = (n - n_{\infty}) / n_{\infty}
$$
where $n$ is the molecules number density of the vapor, $n_{\infty}$ is
the molecule number density of the saturated vapor.

Then the balance of the substance in the closed substance requires
$$
\zeta_* = \zeta + G_1 + G_2
$$
where
$G_1$ is the number of the molecules in the liquid phase around the first
type centers (taken in units of $n_{\infty}$), $G_2$ is the number of
molecules in the droplets formed on the second type centers.

We shall describe the droplet by the value of dimensionless radius $\rho$
which equals to the cube root of the number $\nu$ of the molecules inside
the droplet
$$
\rho = \nu^{1/3}
$$
This value is convenient because the rate of  its growth  is one and the
same for all sizes
$$
\frac{d \rho }{dt} =
\zeta / \tau
$$
where the constant $\tau$ is the characteristic time.
The last equation is valid
for the supercritical droplets under the free molecular regime of the
substance exchange.

Then automatically one can see that the distribution $f_i(\rho, t)$ of the
droplets over $\rho$ in the moment $t$ depends only on the intensity of
the droplets formation in the time when the droplet of given size were
formed.

The application of the ordinary monodisperse approximation
("total monodisperse approximation") is based on
the following explanation:

- Suppose we can suggest an approximation
$$
G_1 \sim  N_{1 tot} \rho_0^3 / n_{\infty}
$$
where
$N_{1 tot}$ is the total number of the droplets formed on the first type
centers during the condensation process.

The last approximation is good when $\rho_0$ is many times grater than
the value of $\rho_0$ at the end of the period of intensive formation
of the droplets on the first type centers, i.e. $\rho_0(\Delta_1 t)$:
$$
\rho_0(t) \gg \rho_0(\Delta_1 t)
$$

This approximation is important when $G_1$ stops the formation of the
droplets on the second type centers, i.e. at $\Delta_2 t$. So, it is necessary
to be
$$
\rho_0(\Delta_1 t) \ll \rho_0(\Delta_2 t)
$$
As far as $d \rho /dt$ is rather smooth function of time one can rewrite
the last estimate as
\begin{equation}\label{ppp}
\Delta_1 t \ll \Delta_2 t
\end{equation}

One can estimate the number $N_{i tot}$ of the droplets formed on the centers
of given sort as
$$
N_{i tot}
\sim
J_{i*} \Delta_i t
$$
where
$$
J_{i*} \equiv f_{i*} n_{\infty}  \tau /  \zeta
$$
is the initial rate of nucleation\footnote{$f_{i*}$ is expressed
in units of $n_{\infty}$}.
Then the violation of (\ref{ppp}) means that
$$
N_{1 tot} \gg N_{2 tot}
$$
As the result the number of the droplets formed on the second type centers
is negligible.

The negligible value of $N_{2 tot}$ in the unique situation when the ordinary
monodisperse approximation fails was the ground to apply this approximation
in the situation of the strong unsymmetry  \cite{Multidecay}.
But we see that if we are interested in the value of $N_{2 tot}$ without
any respect to $N_{1 tot}$ then the  question is still open. Below we shall
resolve this problem.

\section{Special monodisperse approximation}

For $G_i$ one can write the following relation
$$
G_i = \int_0^{\infty}
\rho^3 f_i(\rho, t) d \rho
$$

Now one can analyze the subintegral function
$$
g_i(\rho,t) =
\rho^3 f_i(\rho, t)
$$
which has the sense only for positive $\rho$.

One can see that
$$
g_i(\rho,t) = 0
$$
for
$$
\rho > \rho_0(t) \equiv \int_0^t \frac{\zeta(t')}{\tau} dt'
$$

One can see that
$$
g_i(\rho, t) < \rho^3 f_{i *} \equiv g_{i \ appr}
$$

Consider now the pseudo homogeneous situation. It means that we neglect
the exhaustion of the heterogeneous centers by the droplets.  Here
$$
g_i(\rho, t) \approx \rho^3 f_{i *}
$$
for
all $\rho$
from $$
\rho_0 - \rho < (0.7 \div 0.8) \zeta_* \Delta_i t / \tau $$

As far as
$ \rho^3 f_{i *}$ is the sharp function of $\rho$ we see that
$g_i$ is even more sharp function of $\rho$.  Then it is quite reasonable
to suggest
the monodisperse approximation for $g_i$.

To construct the monodisperse approximation for $g_i$ one has to solve
how to cut the tail at small $\rho$. Despite the rapid decrease at small
$\rho$ the tail can not be integrated at least on the base of approximation
$g_{i \ appr}$ (if we forget about the restriction
$\rho > 0$). There are  two ways to do it.

The first way is to cut off the spectrum on the halfwidth of $g_{i\ appr}$.
It gives the halfwidht
$$\Delta \rho = (1 - 2^{-1/3}) \rho_0 (t)
= 0.21 \rho_0(t) \equiv \Delta_{diff} \rho $$
 which is small in comparison with $\rho_0(g)$. So, really,
the approximation $g_{i\ appr}$ can be used here.

The second way is more close to the iteration procedure from \cite{Kunidec}.
One can define $\Delta \rho$ by the integral way. One can integrate the
approximation $g_{i\ appr}$ from $\rho_0$ up to $0$ taking into account
that $g_i= 0$ for $\rho < 0$. Then
$$
\Delta \rho = \rho_0(t) / 4 \equiv \Delta_{int} \rho$$
As far as $\Delta_{diff} \rho \approx \Delta_{int} \rho$ one can use both
these two ways. The integral way is more convenient as far as it gives
precise asymptotes.

Now we can suggest approximation
$$
G_1 = f_{1*} \Delta \rho \rho_0(t)^3
$$
and rewrite it as
$$
G_1 =  N_{1}(t/4) \rho_0(t)^3
$$
where
$N_{1}(t/4)$ is the number of the droplets formed until the moment $t/4$
and the behavior of $G$ is analyzed in the current moment $t$.

When we consider  the  heterogeneous  condensation  with  essential
exhausting
of centers one can easily note that the function $g_{i}$ becomes even
more sharp. So, the previous derivation is suitable here. Certainly, the
value  of  $N_{1}(t/4)$  has  to  be  calculated  with  account  of
exhaustion of
the heterogeneous centers as it was described in \cite{Multidecay},
\cite{book1}.

Here we can present the last approximation as
$$
G_1 \sim (\eta_{1 tot} - \eta_{1} (t/4)) \rho_0(t)^3
$$
where $\eta_{1}$ is the number of the free heterogeneous centers of the
first sort.

\section{Floating monodisperse approximation}

In \cite{Multidecay} we were interested in the
final parameters of the whole nucleation
periods and used the monodisperse approximation at $t= \Delta_{2} t$. It
allowed to use for $N_{1}(\Delta_{2} t/4)$ the following approximation
$$
N_1(\Delta_{2} t/4) =
\eta_{1 tot} (1- \exp(- B \Delta_{2} t /4) )
$$
where
$$
B = f_{1*} n_{\infty} \zeta_* / \tau \eta_{1 tot}
$$

Now we don't want to use the monodisperse approximation only at $t=\Delta_2
t$. So, we shall act without the last approximation. But now we
have to  use
the monodisperse approximation at the arbitrary moment of time $t$.

We shall use the variables $x,z$ (see \cite{Kunidec}, \cite{Multidecay}) and
shall investigate the system of equations
$$
\zeta_* = \zeta + G_1 + G_2
$$
$$
G_1 = f_{1*} \int_0^z (z-x)^3  \exp(\Gamma_1 (\zeta - \zeta_*)/ \zeta_*)
\theta_1 (x)   dx
$$
$$
G_2 = f_{2*} \int_0^z (z-x)^3  \exp(\Gamma_2 (\zeta - \zeta_*)/ \zeta_*)
\theta_2 (x)   dx
$$
$$
\theta_1 = \exp(- \frac{f_{1*} n_{\infty}}{\eta_{1 tot}}
\int_0^z  \exp(\Gamma_1 (\zeta - \zeta_*)/ \zeta_*)  dx )
$$
$$
\theta_2 = \exp(- \frac{f_{2*} n_{\infty}}{\eta_{2 tot}}
\int_0^z  \exp(\Gamma_2 (\zeta - \zeta_*)/ \zeta_*)  dx )
$$
where $\Gamma_i$ are some parameters (see \cite{Multidecay}), $\theta_i$
are the relative numbers of the free heterogeneous centers of the given
sort.

In the situation of the strong unsymmetry we can rewrite this system in
the following manner
$$
G_1 = f_{1*} \int_0^z (z-x)^3  \exp( - \Gamma_1 G_1 (x) / \zeta_*)
\theta_1 (x)   dx
$$
$$
G_2 = f_{2*} \int_0^z (z-x)^3  \exp( - \Gamma_2 (G_1 + G_2 )/ \zeta_*)
\theta_2 (x)   dx
$$
$$
\theta_1 = \exp(- \frac{f_{1*} n_{\infty}}{\eta_{1 tot}}
\int_0^z  \exp( - \Gamma_1 G_1 (x)/ \zeta_*)  dx )
$$
$$
\theta_2 = \exp(- \frac{f_{2*} n_{\infty}}{\eta_{2 tot}}
\int_0^z  \exp(- \Gamma_2 ( G_1 + G_2 )/ \zeta_*)  dx )
$$

The first and the third equations of the previous system form the closed
system which allows to consider $G_1$ in the second and the forth equations
as some known value. According to the monodisperse approximation it can
be presented as
$$
G_1 (z) =  \frac{f_{1 *}}{E} (1 - \theta_{1}(z/4)) z^3
$$
where
$$
\theta_{1}(z/4) =   \exp(-E \int_0^{z/4} \exp(- \Gamma_1 f_{1*}
 x^4 / 4 \zeta_*) dx )
$$
$$
E = \frac{f_{1*} n_{\infty}}{ \eta_{1 \ tot}}
$$

One can simplify the last expression as
\begin{equation}\label{ap1}
\eta_{1}(z/4) = \eta_{tot}  \exp(-E z/4 )
\end{equation}
for
$$
z/4< z_m
$$
and
\begin{equation}\label{ap2}
\eta_{1}(z/4) = \eta_{tot}  \exp(-E
(\frac{4 \zeta_*}{ \Gamma_1 f_{1*}} )^{1/4}
 A )
\end{equation}
for
$$
z/4 > z_m
$$
where
$$
A = \int_0^{\infty} \exp(-x^4) dx = 0.905
$$
and
$$
z_m =
(\frac{4 \zeta_*}{ \Gamma_1 f_{1*}} )^{1/4}
 A
$$

Then the nucleation on the second sort centers can be described by
the following equations
$$
G_2 = f_{2*} \int_0^z (z-x)^3  \exp( - \Gamma_2
(
\frac{f_{1*}}{E}
(1 - \theta_{1}(z/4)) z^3 + G_2 )/ \zeta_*)
\theta_2 (x)   dx
$$
$$
\theta_2 = \exp(- \frac{f_{2*} n_{\infty}}{\eta_{2 tot}}
\int_0^z  \exp(- \Gamma_2
(
\frac{f_{1*}}{E}
(1 - \theta_{1}(x/4)) x^3 +
G_2)/ \zeta_*)  dx )
$$

We can adopt with a rather high accuracy the following expression for
$\theta_2$ (the reasons are the same as in the section "Final iterations"
in \cite{Multidecay})
$$
\theta_2 = \exp(- \frac{f_{2*} n_{\infty}}{\eta_{2 tot}}
\int_0^z  \exp(- \Gamma_2
(
\frac{f_{1*}}{E}
( 1 - \theta_{1}(x/4)) x^3 +
f_{2*} x^4/4
)/ \zeta_*)  dx )
$$
or with the help of approximation
(\ref{ap1}), ({\ref{ap2}) it can be presented in more simple form.

If
$z/4 < z_m$ then
\begin{equation} \label{pp}
\theta_2 = \exp(- \frac{f_{2*} n_{\infty}}{\eta_{2 tot}}
\int_0^z  \exp(- \Gamma_2
(
\frac{f_{1*}}{E}
(1  -
   \exp(-E x/4 )
) x^3 +
f_{2*} x^4/4
)/ \zeta_*)  dx )
\end{equation}

If
$z/4>z_m$ then
\begin{eqnarray} \label{pwpp}
\theta_2 = \exp(- \frac{f_{2*} n_{\infty}}{\eta_{2 tot}} (
\int_0^{4z_m}  \exp(- \Gamma_2
(
\frac{f_{1*}}{E}
(1 -    \exp(-E x/4 )) x^3 +
f_{2*} x^4/4
)/ \zeta_*)  dx
+ \nonumber \\  \\ \nonumber
\int_{4z_m}^z  \exp(- \Gamma_2
(
\frac{f_{1*}}{E}
( 1  -    \exp(-E
(\frac{4 \zeta_*}{ \Gamma_1 f_{1*}} )^{1/4}
 A )           ) x^3 +
f_{2*} x^4/4
)/ \zeta_*)  dx )
)
\end{eqnarray}

Now we have to calculate the integrals appeared in the last two expressions.
We shall start from the first one.

Consider (\ref{pp}).
One can see that function
$$
  \phi \equiv
\Gamma_2
(
\frac{f_{1*}}{E}
( 1 -
   \exp(-B x/4 )
) x^3 +
f_{2*} x^4/4
)/ \zeta_*
$$
is very sharp function. It is more sharp than
$$
  \phi_0 \equiv
const x^3 + const x^4
$$

The function
$ (1 -
  \exp(-B x/4 )
)$
is rather smooth in comparison with $\phi$ and $\phi_0$.

One can note that integrals
$\int_0^{\infty} \exp(-x^3 )dx  = 0.89 $ and
$\int_0^{\infty} \exp(-x^4)dx  = 0.90 $
are approximately equal. Both subintegral  functions have very sharp  back
front and one can speak about the cut-off in both cases.
 The approximate equality  of these integrals
means that these cut-off have approximately same values.

We shall define the characteristic parameter $z_q$ by equality
\begin{equation}
 \Gamma_2
(
\frac{f_{1*}}{E}
(1 -
   \exp(-E z_q/4 )
) z_q^3 +
f_{2*} z_q^4/4
)/ \zeta_*
=1
\end{equation}

Then the integral in (\ref{pp}) can be rewritten as
\begin{eqnarray}
\int_0^z  \exp(- \Gamma_2
(
\frac{f_{1*}}{E}
( 1 -
   \exp(-E x/4 )
) x^3 + \nonumber \\ \\ \nonumber
f_{2*} x^4/4
)/ \zeta_*)  dx  = \Theta(z- C z_q) C z_q +
\Theta( C z_q - z)  z
\end{eqnarray}
where
$$
C = \frac{1}{2} (\int_0^{\infty} \exp(-x^4) dx +
\int_0^{\infty} \exp(-x^3) dx )
$$

This representation of the integral transfers (\ref{pp}) into
\begin{equation} \label{pp1}
\theta_2 = \exp(- \frac{f_{2*} n_{\infty}}{\eta_{2 tot}}
(
\Theta(z- C z_q) C z_q +
\Theta( C z_q - z)  z)
)
\end{equation}

Now we shall analyze the integral in  (\ref{pwpp}). The reasons are the
same. We shall introduce parameter $z_l$ by the following relation
$$
           \Gamma_2
(
\frac{f_{1*}}{E}
 ( 1  -    \exp(-E
(\frac{4 \zeta_*}{ \Gamma_1 f_{1*}} )^{1/4}
 A )           ) z_l^3 +
f_{2*} z_l^4/4
)/ \zeta_*
 = 1
$$
Note that we need only one parameter as far as
$$
  1 -    \exp(-E x/4 )    \leq
                 1 -   \exp(-E
(\frac{4 \zeta_*}{ \Gamma_1 f_{1*}} )^{1/4}
 A )
$$
for $\frac{x}{4} < z_m$
and
$$
 1 -    \exp(-E x/4 )  |_{x/4=z_m}   \approx
1 -    \exp(-E
(\frac{4 \zeta_*}{ \Gamma_1 f_{1*}} )^{1/4}
 A )
$$

If $z_l   <  4 z_m$ then
\begin{eqnarray}
\int_0^{4z_m}  \exp(- \Gamma_2
( \frac{f_{1*}}{E} ( 1 -    \exp(-E x/4 )) x^3 +
f_{2*} x^4/4
)/ \zeta_*)  dx
 \gg \nonumber \\ \\ \nonumber
\int_{4z_m}^z  \exp(- \Gamma_2
(
\frac{f_{1*}}{E}
( 1 -    \exp(-E
(\frac{4 \zeta_*}{ \Gamma_1 f_{1*}} )^{1/4}
 A )           ) x^3 +
f_{2*} x^4/4
)/ \zeta_*)  dx
\end{eqnarray}
and one can analyze only
$$
I_1 =
\int_0^{4z_m}  \exp(- \Gamma_2
(
\frac{F_{1*}}{E}
(1  -    \exp(-E x/4 )) x^3 +
f_{2*} x^4/4
)/ \zeta_*)  dx
$$
It was already done in consideration of (\ref{pp}).

If $z_l   >  4 z_m$ then both
$$
I_1 =
\int_0^{4z_m}  \exp(- \Gamma_2
(
\frac{f_{1*}}{E}
( 1 -    \exp(-E x/4 )) x^3 +
f_{2*} x^4/4
)/ \zeta_*)  dx
$$
and
$$
I_2 =
\int_{4z_m}^z  \exp(- \Gamma_2
(
\frac{f_{1*}}{E}
( 1 -    \exp(-E
(\frac{4 \zeta_*}{ \Gamma_1 f_{1*}} )^{1/4}
 A )           ) x^3 +
f_{2*} x^4/4
)/ \zeta_*)  dx
$$
are essential.

Then
$$
I_1 = 4 z_m
$$
and $I_2$ can be analyzed quite analogously. Namely, we shall introduce
$z_t$  from equality\footnote{Certainly $ z_t = z_l $ }
$$                    \Gamma_2
(
\frac{f_{1*}}{E}
 ( 1 -   \exp(-E
(\frac{4 \zeta_*}{ \Gamma_1 f_{1*}} )^{1/4}
 A )           ) z_t^3 +
f_{2*} z_t^4/4
)/ \zeta_* =1
$$

If $z_t$ is near $4 z_m$ then
$I_2
$ is small in comparison with $I_1$ and there is no need to analyze $I_2$.

If $I_2$ is essential in comparison with $I_1$ one can use the following
approximation:
$$
I_2 =
( z- 4 z_m )
$$
for
$z< C z_t$
$$
I_2 = C z_t - 4 z_m
$$
for
$ z > C z_t$.

This completes the approximate analysis of the expression for $\theta_2$.

Some parameters $z_l$, $z_m$, $z_q$, $z_t$ may coincide but they are conserved
in order to avoid  misunderstanding.

The main interesting value is $\theta_2(\infty)$. The final expressions
for this value are more simple. They can be directly obtained from the
already presented ones.

On the base of $\theta_{i}(z)$ one can easily find the number of droplets
$N_i$ as
$$
N_i = \eta_{tot\ i}(1  - \theta_i )
$$
To find the total number of the droplets one has to put the arguments
to $\infty$.

\end{document}